\documentclass[nobalancelastpage,twocolumn,floatfix,superscriptaddress,aps,prx]{revtex4}

\usepackage{microtype}
\usepackage[hidelinks,colorlinks,citecolor=blue,linkcolor=black,urlcolor=blue]{hyperref}
\usepackage{graphicx}
\usepackage{bm}
\usepackage{bbm}
\usepackage[bb=boondox]{mathalfa}
\usepackage{nicefrac}
\usepackage[caption=false]{subfig}
\usepackage{verbatim}
\usepackage{scrextend}
\usepackage{tikz}
\usepackage{bbold}

\usepackage{amsfonts}
\usepackage{amssymb}
\usepackage{amsmath}
\usepackage{enumerate}
\usepackage{multirow}
\usepackage{microtype}
\usepackage{mhchem}
\usepackage{siunitx}
\DeclareSIUnit\gauss{G}
\usepackage[capitalise]{cleveref}
\usepackage{tikz}
\usepackage{chngcntr}

\usetikzlibrary{positioning}
\usetikzlibrary{fit}
\usetikzlibrary{calc,shadings}
\usetikzlibrary{patterns}

\newcommand{\ket}[1]{| #1 \rangle}
\newcommand{\kket}[1]{| #1 \rangle\hspace{-2.5pt}\rangle}
\newcommand{\bra}[1]{\langle #1 |}

\newcommand{\oouter}[1]{| #1 \rangle\hspace{-2.5pt}\rangle\langle\hspace{-2.5pt}\langle #1 |}

\begin{document}
\title{Dissipative discrete time crystals}

\author{James O'Sullivan}
\affiliation{London Centre for Nanotechnology, University College London, 17-19 Gordon Street, London, WC1H 0AH, UK}

\author{Oliver Lunt}
\affiliation{Department of Physics, University College London, Gower Street, London, WC1E 6BT}

\author{Christoph W. Zollitsch}
\affiliation{London Centre for Nanotechnology, University College London, 17-19 Gordon Street, London, WC1H 0AH, UK}

\author{M. L. W. Thewalt}
\affiliation{Department of Physics, Simon Fraser University, Burnaby, British Columbia, Canada}

\author{John J. L. Morton}
\affiliation{London Centre for Nanotechnology, University College London, 17-19 Gordon Street, London, WC1H 0AH, UK}

\author{Arijeet Pal}
\email{a.pal@ucl.ac.uk}
\affiliation{Department of Physics, University College London, Gower Street, London, WC1E 6BT}
\affiliation{Rudolf Peierls Centre for Theoretical Physics, Oxford University, Oxford OX1 3PU, UK}
\date{\today}

\begin{abstract}
Periodically driven quantum systems host a range of non-equilibrium phenomena which are unrealizable at equilibrium. Discrete time-translational symmetry in a periodically driven many-body system can be spontaneously broken to form a discrete time crystal, a putative quantum phase of matter. We present the observation of discrete time crystalline order in a driven system of paramagnetic \ce{P}-donor impurities in isotopically enriched \ce{^{28}Si} cooled below \SI{10}{\kelvin}. The observations exhibit a stable subharmonic peak at half the drive frequency which remains pinned even in the presence of pulse error, a signature of DTC order. We propose a theoretical model based on the paradigmatic central spin model which is in good agreement with experimental observations, and investigate the role of dissipation in the stabilisation of the DTC. Both experiment and theory indicate that the order in this system is primarily a dissipative effect, and which persists in the presence of spin-spin interactions. We present a theoretical phase diagram as a function of interactions and dissipation for the central spin model which is consistent with the experiments. This opens up questions about the interplay of coherent interaction and dissipation for time-translation symmetry breaking in many-body Floquet systems. 
\end{abstract}

\maketitle

\section{Introduction}

\begin{figure*}[ht]
    \centering
    \subfloat[]{
        \centering
        \includegraphics[width=0.45\linewidth]{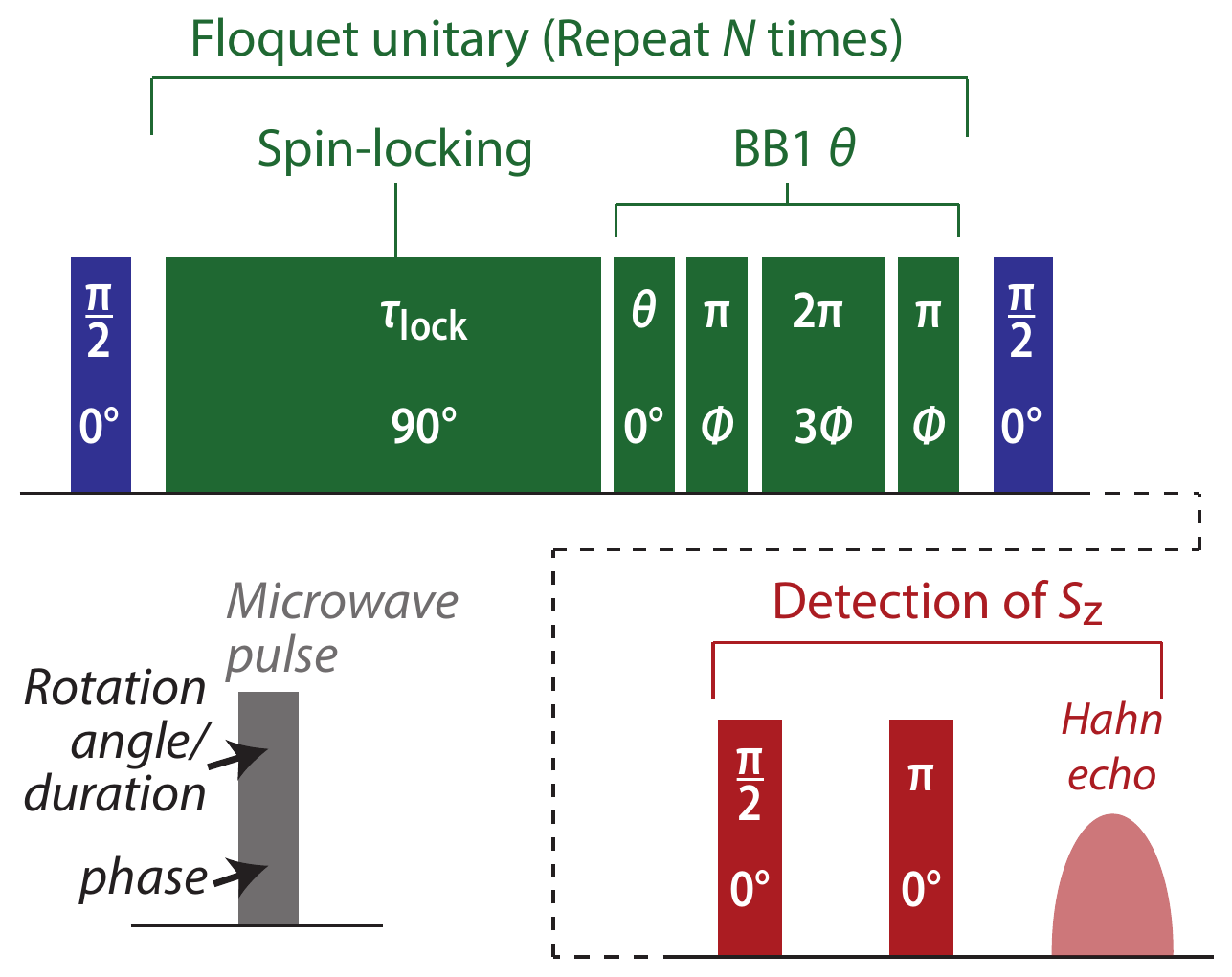}
        \label{fig:pulse_sequence}
    }
	\subfloat[]{
		\centering
		\includegraphics[width=0.45\linewidth]{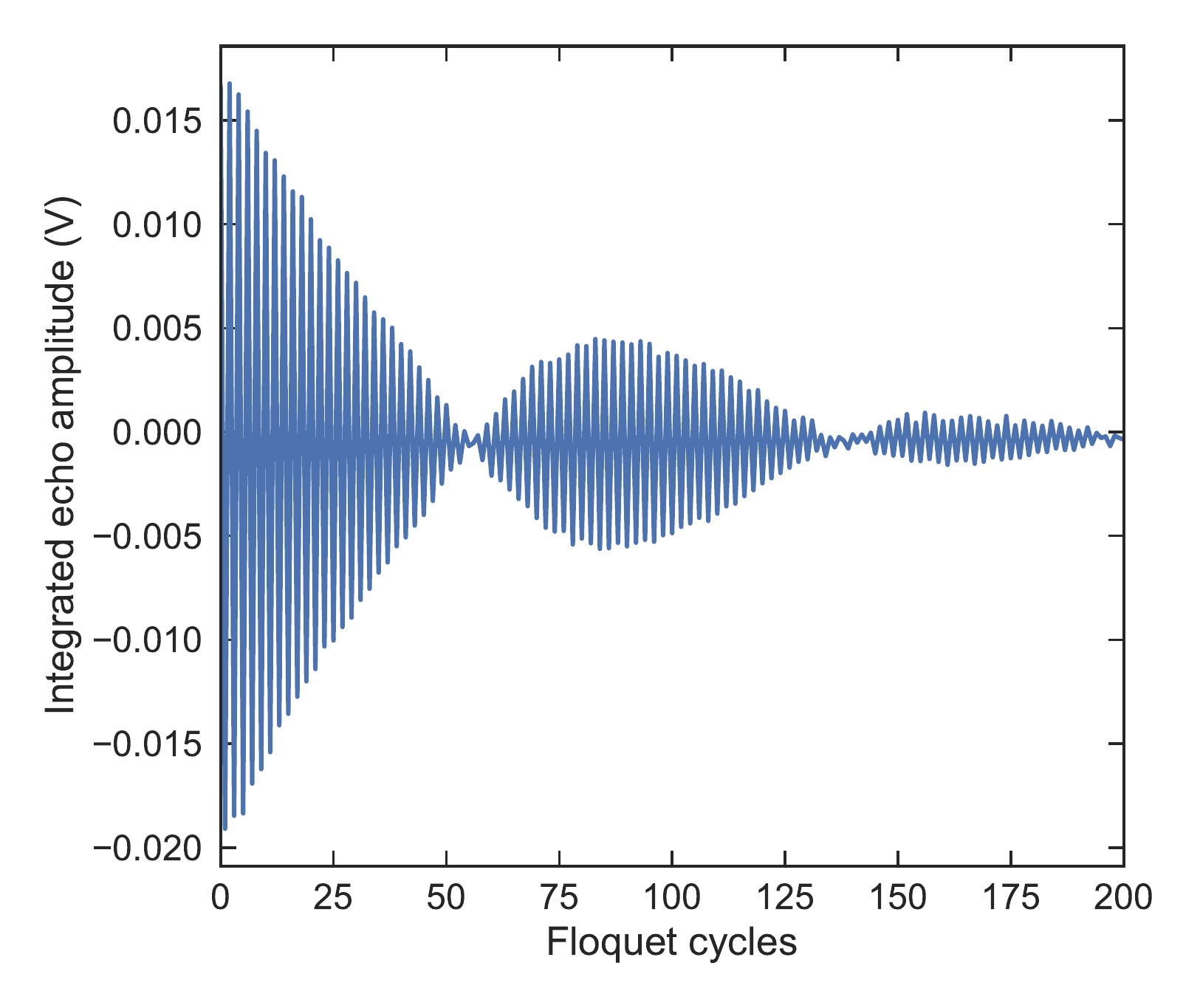}
		\label{fig:pulse}
	}
    \vspace{-0.5em}
    
    \subfloat[]{
        \centering
        \includegraphics[width=0.45\linewidth]{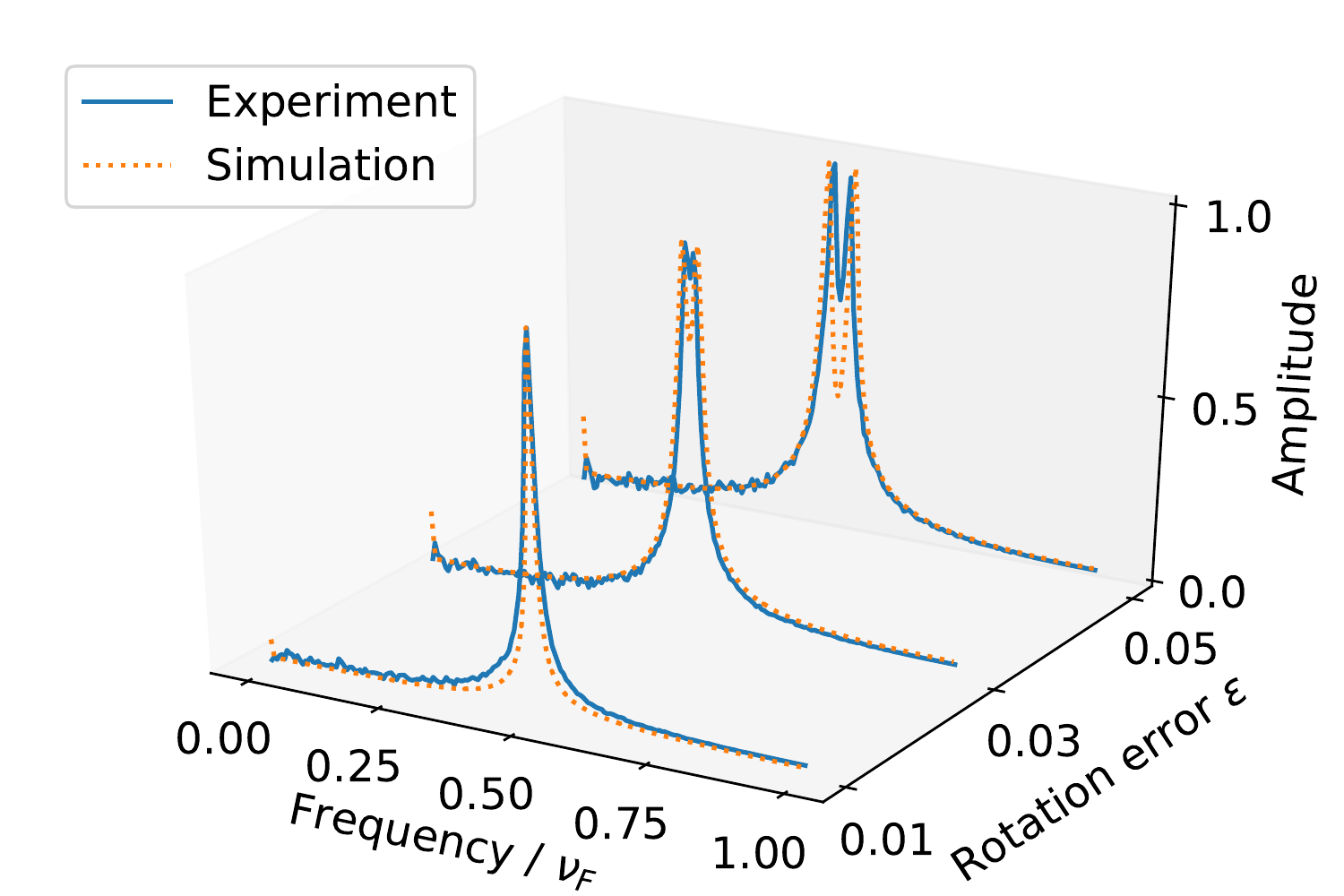}
        \label{fig:3D_exp_vs_theory_comparison}
    }
    \subfloat[]{
        \centering
        \includegraphics[width=0.45\linewidth]{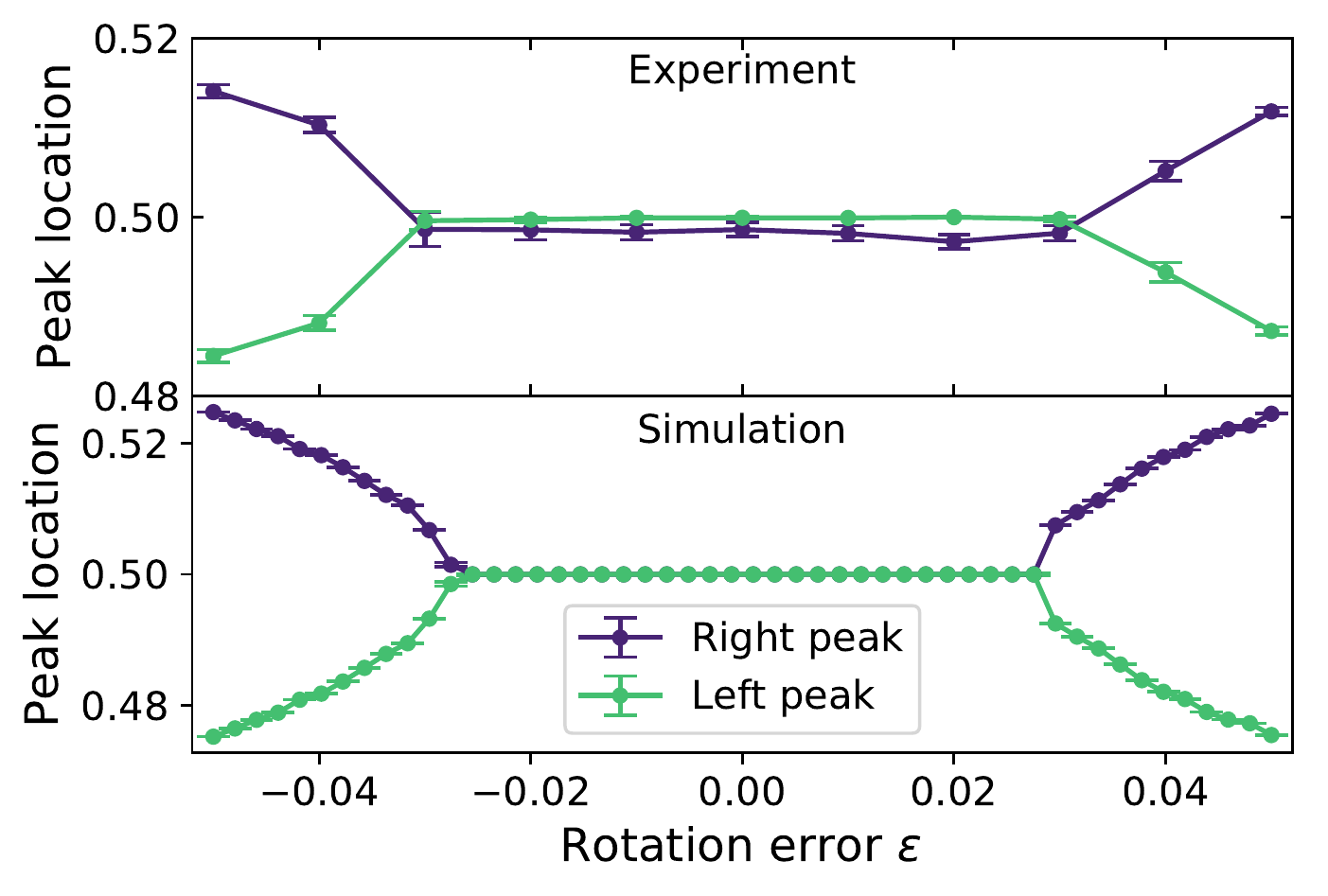}
    }
    
\caption{Microwave pulse control sequence and observation of the time crystal phase. \textbf{(a)} We initialise the spin ensemble with a $\pi/2$-pulse, then begin driving with $N$ Floquet unitaries, before applying a final $\pi/2$-pulse and read out with a $\pi/2-\pi$ Hahn echo sequence. \textbf{(b)} Oscillating signal as we sweep the number of Floquet cycles from 0 to 200, with $\tau_{\mathrm{lock}}=10\pi, \theta=0.99\pi$. This signal can be Fourier transformed to analyse the strength of the $\nu = 1/2$ oscillations. \textbf{(c)} A comparison between experiment and theory of the Fourier spectrum of the net magnetization as a function of  rotation error $\epsilon$, calculated over 200 Floquet cycles and vertically scaled to 1 (see \cref{sec:model} for details of the theoretical model). For nonzero $\epsilon \sim 0.01$, the Fourier peak remains unsplit, indicating the presence of the DTC phase. For larger values of $\epsilon$, we return to the trivial symmetry-unbroken phase. \textbf{(d)} Comparison of the Fourier peak location as a function of rotation error $\epsilon$, using a double Lorentzian fit. There is a finite window around $\epsilon = 0$ where the peak remains rigidly locked at half the drive frequency, indicating the robustness of the DTC phase. \textbf{Parameters (c,\,d):} $\kappa^{1} = \SI{2.3}{\kilo\Hz}$, $\kappa^{2} = \SI{0.9}{\kilo\Hz}$, $J = \SI{300}{\Hz}$, $h = \SI{10}{\Hz}$, $N = 8$, $t_{\pi} = \SI{300}{\nano\second}$, $\tau_{\mathrm{lock}} = 100 t_{\pi}$.}

\end{figure*} 

A central paradigm in condensed matter physics has been to classify phases of matter by their symmetries. Indeed, spontaneous symmetry breaking describes many known phase transitions. A common  symmetry-breaking phase is a crystal in real space, where the symmetry under continuous spatial translation is broken to a lower discrete one. A natural question is then to ask whether it is possible to analogously break \textit{time}-translation symmetry \cite{Wilczek_TimeCrystal2012}. Breaking of continuous time-translation symmetry has been shown to be impossible in thermal equilibrium \cite{Bruno_NoTC2013, Watanabe_AbsenceTC2015}, but periodically-driven (`Floquet') systems provide the means to break discrete time-translation symmetry, thereby forming a discrete time crystal \cite{Sacha_TC2015, Khemani_PiSpinGlass2016, Else_DTC2016, Keyserlingk_PhaseII_2016, Yao_DTC2017, moessner2017_QFloquetMatter, sacha2017time}. Observations consistent with discrete time-crystalline behaviour were reported soon after in experiments \cite{Choi2017, zhang2017_DTC, Rovny_DTC2018, Pal_NMRDTC2018}. 

In this article we report the observation of a discrete time crystal (DTC) in silicon doped with phosphorus. Silicon provides an ideal platform for implementing the dynamic pulse sequences crucial for realizing time crystals, owing to its ability to be isotopically engineered to an exceptionally high purity \cite{Becker2010}, having the longest coherence times of any solid state system \cite{Tyryshkin2012,Wolfowicz2013}, and its versatility with dopant species and concentration. 

In phosphorus-doped silicon, the dopant electron spins interact via dipolar interactions. At donor concentrations below about $10^{16}$cm$^{-3}$ these interactions are weak compared to dissipative effcts due to magnetic impurities and charge noise. Therefore the DTC order in this system is primarily produced by \textit{dissipation}. Dissipation and nonlinearities provide a natural route to produce robust period-doubling, as has been observed with AC-driven charge density waves \cite{brownSubharmonicShapiroSteps1984,wiesenfeldNoiseToleranceFrequencylocked1987,balentsTemporalOrderDirty1995} and Faraday wave instabilities \cite{crossPatternFormationOutside1993}. For intuition, one could imagine a dissipative system with two basins of attraction, and a periodic drive which flips between these two attractors. It is clear when the flip is perfect that, starting from one of the attractors, this system will exhibit period-doubling. Even when the flip is not perfect, provided the state is sufficiently close to one of the attractors, the dissipation will cancel out any imperfections and yield robust breaking of time-translation symmetry \cite{Iemini_BTC2018, Gong_DTC2018, gambettaDiscreteTimeCrystals2019}. 

However, in an interacting many-body quantum system, it is not clear whether this dissipation-induced symmetry breaking is stable in the presence of interactions. If the interactions do not preserve the attractors (e.g.\ if the interactions do not commute with the dissipation), then in a large enough system the interactions may have a destabilizing effect and destroy the DTC order.

We present observations of robust DTC order over 200 Floquet periods using a sample of phosphorus-doped silicon in the strong dissipation and weak interactions regime, indicating that this dissipative DTC order is indeed stable to weak interactions. We go on to produce a theoretical phase diagram of a dissipative DTC as a function of dissipation rate and interaction strength, and experimentally probe the dissipative DTC and normal phases. We produced the phase diagram using the driven central-spin model coupled to a dissipative bath, direct simulations of which using experimental parameters agree well with our experimental observations.

As a point of independent interest, in \cref{sec:BB1} we comment on the use of the crystalline fraction as a DTC order parameter. In the experiment we used composite BB1 pulses\cite{Wimperis1994,Morton2005} to enable extremely uniform spin rotations. This is important because in experiments DTC order is usually observed by fixing some nonzero rotation error and probing how robust the DTC order is to this error. One proposed experimental probe is the crystalline fraction \cite{Choi2017,choi2018_DTC_Thermal,Rovny_DTC2018}, which roughly measures the strength of the subharmonic Fourier peak compared with the rest of the spectrum. A `plateau' in the crystalline fraction across a finite window of rotation errors has been used as an indicator of DTC order. However, we argue that this plateau can arise simply from non-uniform spin rotations, since the plateau disappears when we use BB1 pulses to ensure uniform rotations. This indicates that using crystalline fraction as a witness for DTC order can obfuscate the different effects.

We note here that we do not expect this system to be many-body localized (MBL) \cite{basko2006metal, gornyi2005interacting, pal2010mbl, Huse2013LPQO, bahriLocalizationTopologyProtected2015, Pekker2014HG, NandkishoreHuse_review, abanin2018ergodicity,oganesyanLocalizationInteractingFermions2007a,oganesyanEnergyTransportDisordered2009,parameswaranManybodyLocalizationSymmetry2018,aletManybodyLocalizationIntroduction2018}. MBL has been argued to be a necessary condition for the existence of time-crystallinity  \cite{Lazarides_OpenDTC_2017}, but recent observations of a discrete time crystal have been reported in systems which may not be many-body localized \cite{Choi2017, zhang2017_DTC, Rovny_DTC2018, Pal_NMRDTC2018}, due to the presence of a bath, and dipolar interactions \cite{Yao_DipolarMBL_2014, burin2015_longrangeMBL, Levi_MBLdissipation_2016, Medvedyeva_MBLdissipation_2016}, which are long-ranged in 3D. Though it has been suggested that MBL can survive in systems with long-range interactions \cite{nandkishoreManyBodyLocalizationLongRange2017}, we do not expect those arguments to hold here, as they rely upon the Anderson-Higgs mechanism. 

\section{Pulse protocols in doped silicon}
\label{sec:pulse_protocols}
A periodic pulse sequence is implemented on the electronic spins as shown in \cref{fig:pulse_sequence}. We use a sample of \ce{^{28}Si} enriched to 99.995\%; this provides a magnetically clean environment which gives the dopant spins an exceptionally narrow linewidth of less than \SI{10}{\micro\tesla}. The sample is placed in a Bruker X-band microwave resonator and microwave driving is applied at a frequency of \SI{9.77459}{\giga\hertz}. We begin with a spin ensemble polarised in the $+z$ direction, parallel to an applied magnetic field setpoint of \SI{3512}{\gauss}. The system is initialised with a  $\pi/2$-pulse, rotating the spins into the $x-y$ plane. Next, a long microwave `spin-locking' pulse is applied along the $x$ direction for a duration $\tau_\textrm{lock}$, which serves to allow the spins to interact while decoupling them from their environment. We then immediately rotate about the $y$ axis by $\pi$ with a deliberate error $\epsilon$. This rotation is achieved via the use of a composite BB1 pulse\cite{Wimperis1994,Morton2005}, which compensates for variations in the Rabi frequency across the sample. This was crucial to removing artefacts in the data that can otherwise be mistaken for features of a DTC (see \cref{sec:BB1}). Together, the spin-locking and BB1 pulse form a single Floquet unitary. After applying $N$ Floquet unitaries we rotate the ensemble back to the $z$ axis with a $\pi/2$-pulse. We then read out the resulting polarisation using a $\pi/2$-$\pi$ Hahn echo sequence. The time $\tau_\textrm{lock}$ is chosen to correspond to an integer multiple of $2\pi$-rotations of the spins about the $x$ axis to avoid any dynamical decoupling effects.

\begin{figure}[t]
    \centering
    \subfloat[]{
        \centering
        \includegraphics[scale=0.6]{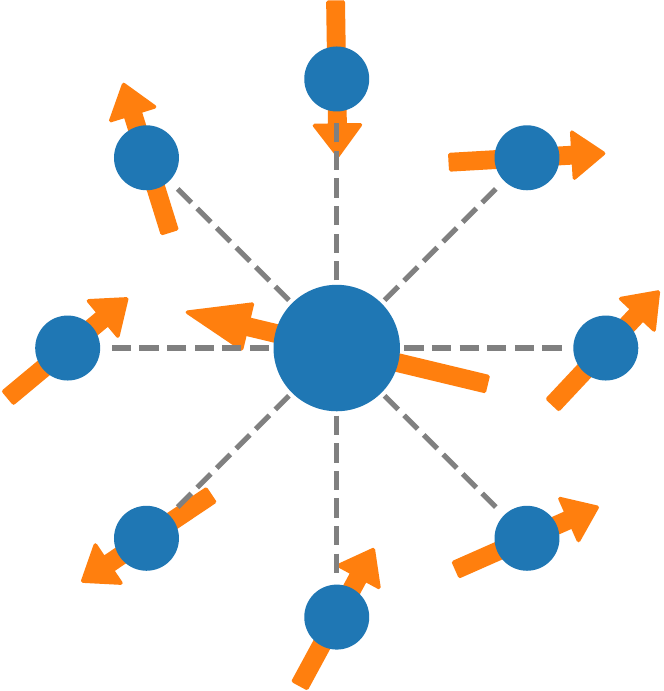}
        \label{fig:central_spin_model_cartoon}
    }
    \subfloat[]{
        \centering
        \includegraphics[scale=0.35]{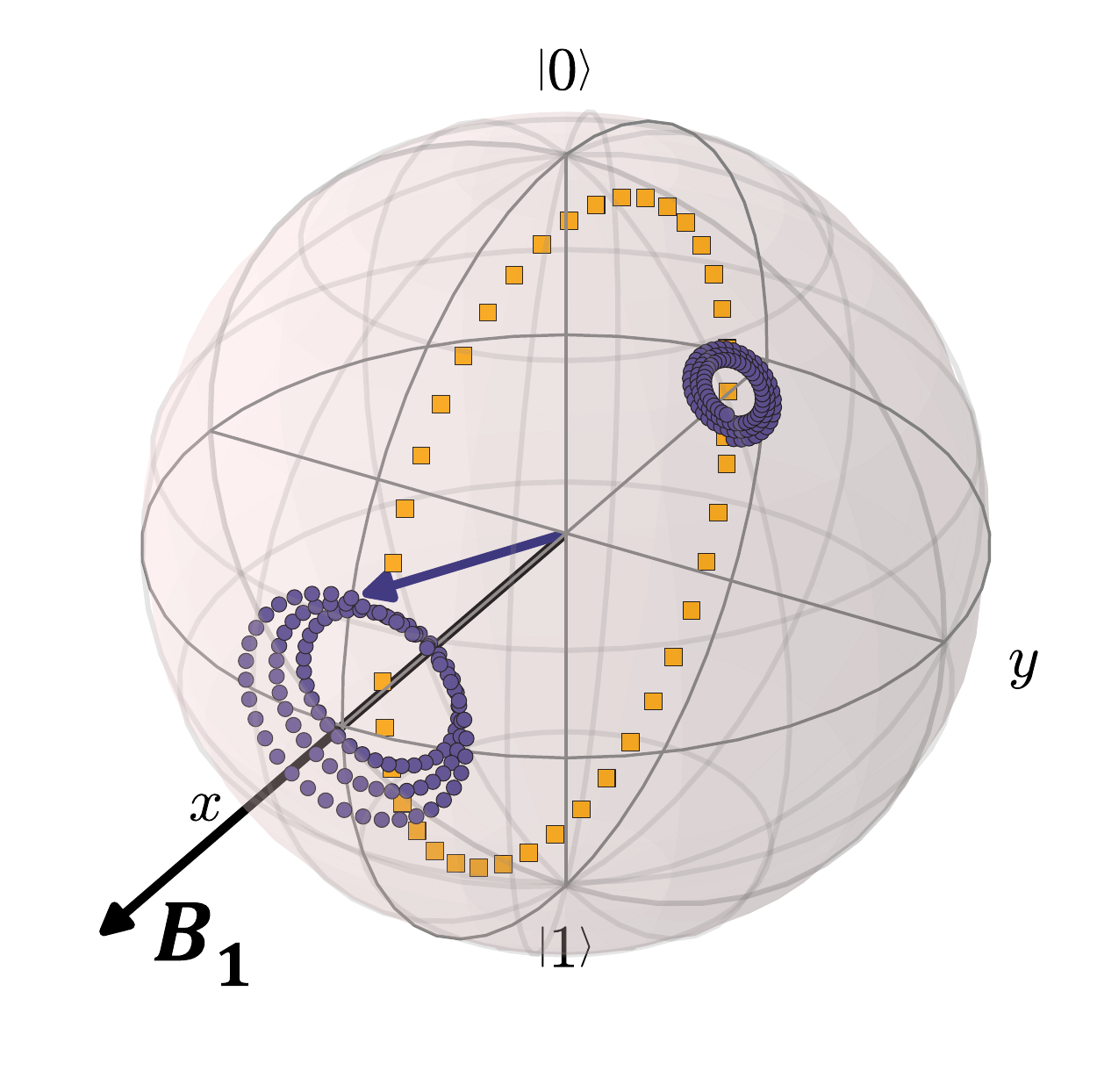}
        \label{fig:bloch_floquet}
    }
    
    \vspace{-1.5em}
    \subfloat[]{
        \centering
        \includegraphics[width=\linewidth]{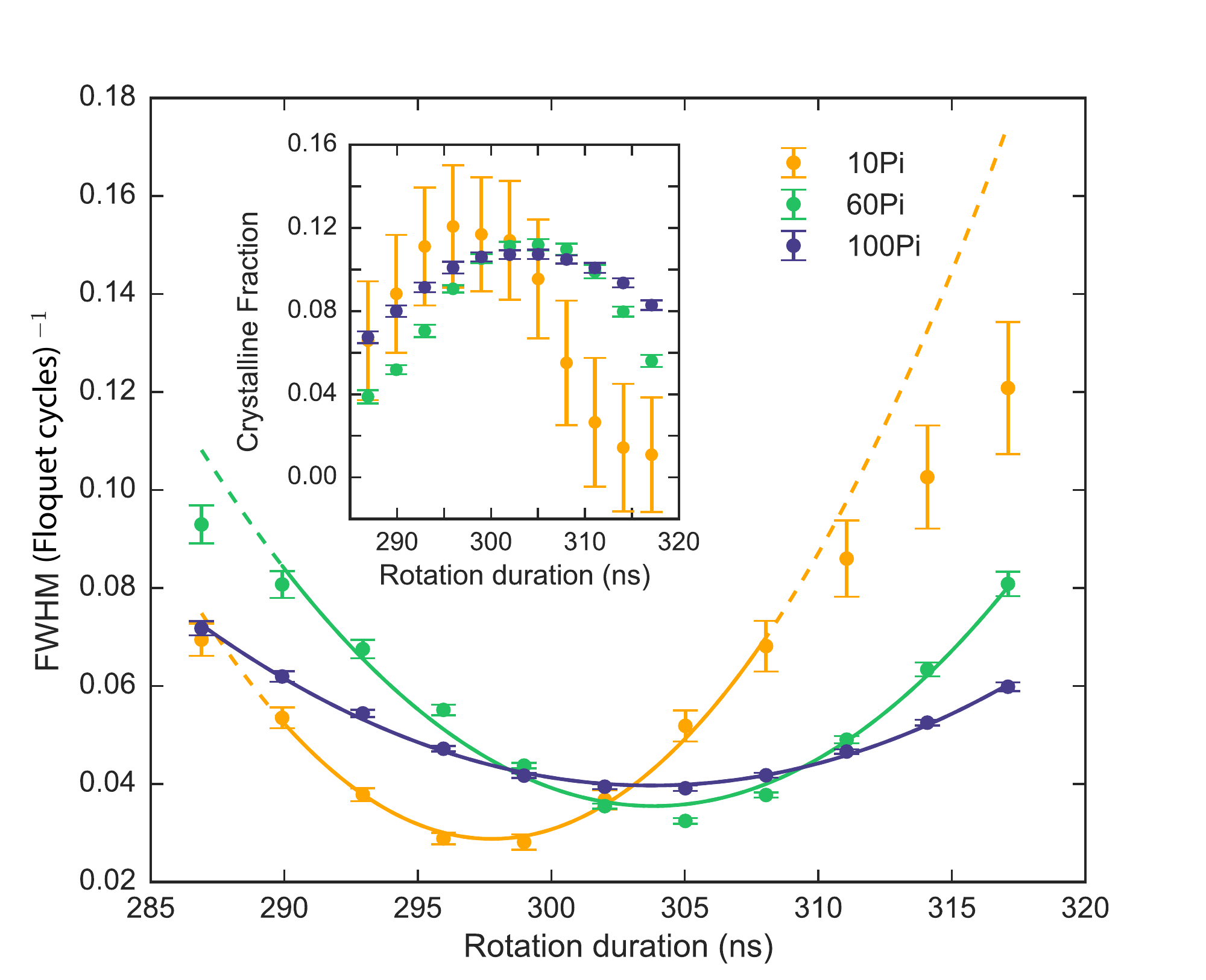}
        \label{fig:experiment_peak_width}
    }
    \caption{\textbf{(a)} A cartoon of the central spin model. \textbf{(b)} A Bloch sphere visualization of the refocusing effect enacted by dephasing. \textbf{(c)} Experimental data of the full width at half maximum of the Fourier peak of the net magnetization as a function of rotation duration, plotted for different interaction times $\tau_{\mathrm{lock}}$. The lines are quadratic fits; solid line region indicates the points that were used for data fitting, beyond which the data began to deviate from a quadratic dependence. The fit indicates that for small rotation errors $\epsilon$ the peak width follows the expected $\epsilon^{2}$ dependence for a dissipative DTC. The rotation time corresponding to a perfect spin-flip varied slightly for different spin-locking times, which is why the bottoms of the parabolas occur at slightly different rotation durations. \textbf{(inset)} Crystalline fraction against rotation duration. This is defined as the ratio of the $\nu = 1/2$ peak to the total spectral power, $f = |S(\nu = 1/2)|^2/\sum_{\nu}|S(\nu)|^2$ \cite{Choi2017}, and is a measure of the total fraction of spins in the DTC phase. The fraction decreases as the rotation duration deviates further from $\pi$ as expected; the rate of decrease is slower for longer spin-locking.}
\end{figure}
\section{Experimental observations}
\label{sec:experimental_observations}
We sweep the the number of Floquet cycles $n$ from 0 to 200 with error $\epsilon$ and map out the integrated echo amplitude as a function of $n$. The result is a decaying oscillating signal with frequency $\nu = 1/2$. We take the Fourier transform of these oscillations and examine what happens to the $\nu = 1/2$ peak as we change $\epsilon$. In the absence of interactions between electron spins, when we increase $\epsilon$ on the $\pi$-pulse in each Floquet cycle we expect the cumulative error caused by successive over- or under-rotation of the spin ensemble to cause modulation of the oscillations. This is apparent in the Fourier transform as a splitting of the $\nu = 1/2$ peak, as shown in \cref{fig:3D_exp_vs_theory_comparison}. Increasing $\epsilon$, we expect to see these peaks split further apart as the error in the $\pi$ rotations accumulates faster leading to faster modulation. However, in the presence of sufficiently strong dissipation, we do not see an immediate splitting of the $\nu = 1/2$ peak, but instead find a region where the oscillations remain resilient to error. 

We use a sample of \ce{^{28}Si} doped with \SI{1e15}{\cm^{-3}} phosphorous. The phase coherence time $T_2$ was measured as \SI{260}{\micro\second}, spin-lattice relaxation time $T_1$ was measured as \SI{1097}{\micro\second} and the Rabi frequency was \SI{1.650}{\mega\hertz}, corresponding to a $\pi$-pulse duration of \SI{303}{\nano\second}. The decay rate of the driven oscillations $T_{1 \rho}$ was measured at \SI{190}{\micro\second} from an exponential fit. Further, in \cref{fig:high_conc} we investigate another sample doped with \SI{3e15}{\cm^{-3}} \ce{P} in \ce{^{28}Si} and show that despite a 200\% increase in spin concentration, the key features remain the same, indicating that we are still in the dephasing dominated regime.

We can tune the spin-locking time $\tau_\textrm{lock}$. Increasing $\tau_\textrm{lock}$ increases both the dissipation and interaction time per Floquet cycle of the spins. For sufficiently long $\tau_\textrm{lock}$ we observe a range of values of $\epsilon$ for which the central $\nu = 1/2$ peak does not split. Further increasing  $\tau_\textrm{lock}$ increases the range of $\epsilon$ for which we have a single peak, indicating that the oscillations have increased resilience to errors.


One indication that the DTC order in this experimental system is effectively described by a dephasing model can be found by looking at the dependence of the Fourier peak width on the rotation error $\epsilon$. In the strong-dephasing limit, analysis of a simple single-spin model indicates that the XY-dephasing results in exponential decay of $\left|\langle \sigma^{z} \rangle \right|$ at a rate $\Gamma \sim -\log[\cos\epsilon\pi] \sim (\epsilon\pi)^{2}/2$ per Floquet period, which is quadratic in $\epsilon$ for $\epsilon \ll 1$ (see \cref{sec:derivation}). \cref{fig:experiment_peak_width} shows the Fourier peak width as a function of $\epsilon$ for the experimental data, which shows the predicted $\epsilon^{2}$ dependence. There are additional contributions to the peak width from other sources, such as errors from finite rotation times, but the $\epsilon^{2}$ contribution can be seen as a hallmark of the \textit{dissipative} DTC phase.

\section{Phase diagram}
\label{sec:model}

\begin{figure*}[htp]
    \centering
    \subfloat[]{
        \centering
        \includegraphics[width=0.5\linewidth]{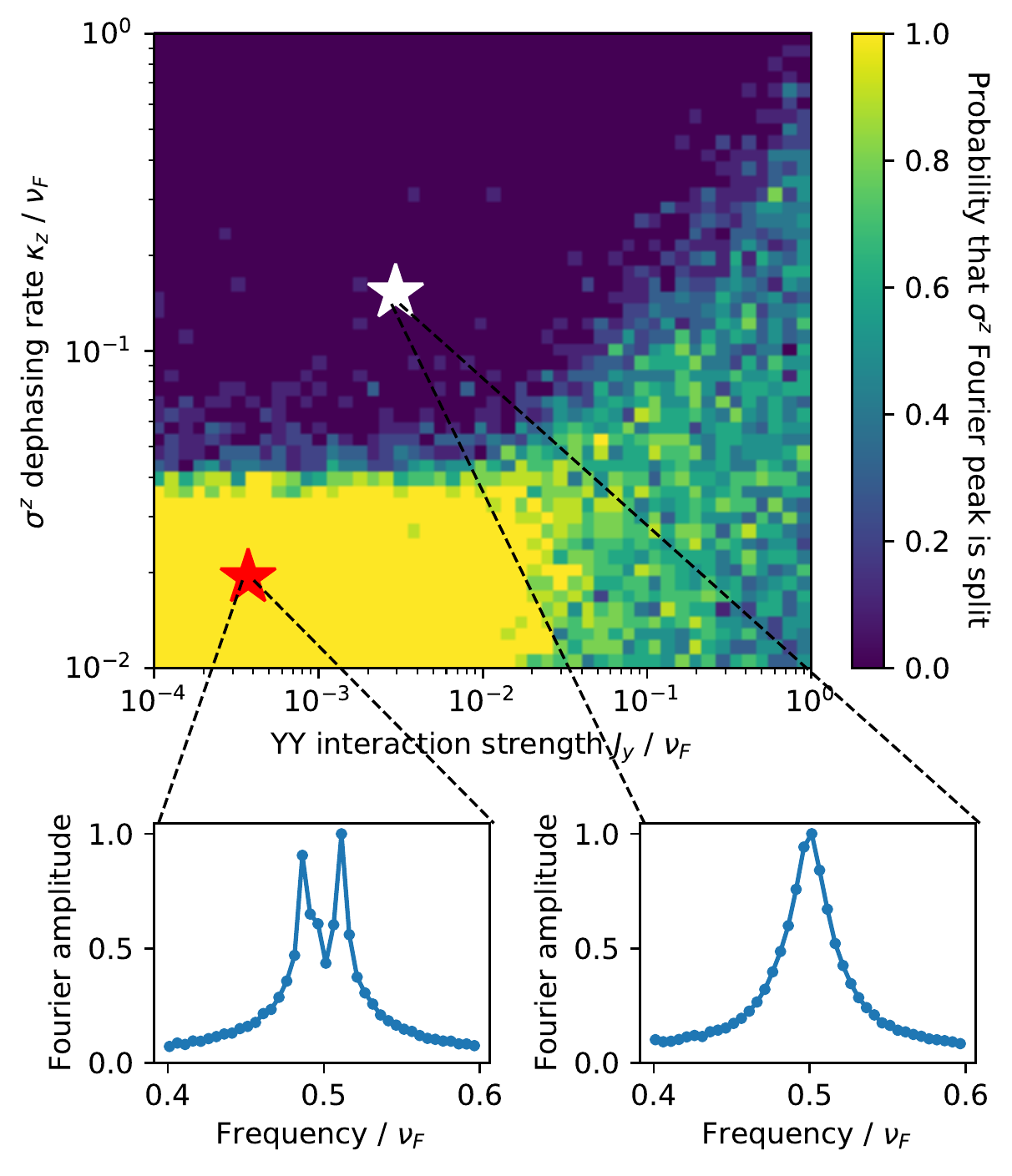}
        \label{fig:sigmaz_phase_diagram}
    }
    \subfloat[]{
        \centering
        \includegraphics[width=0.5\linewidth]{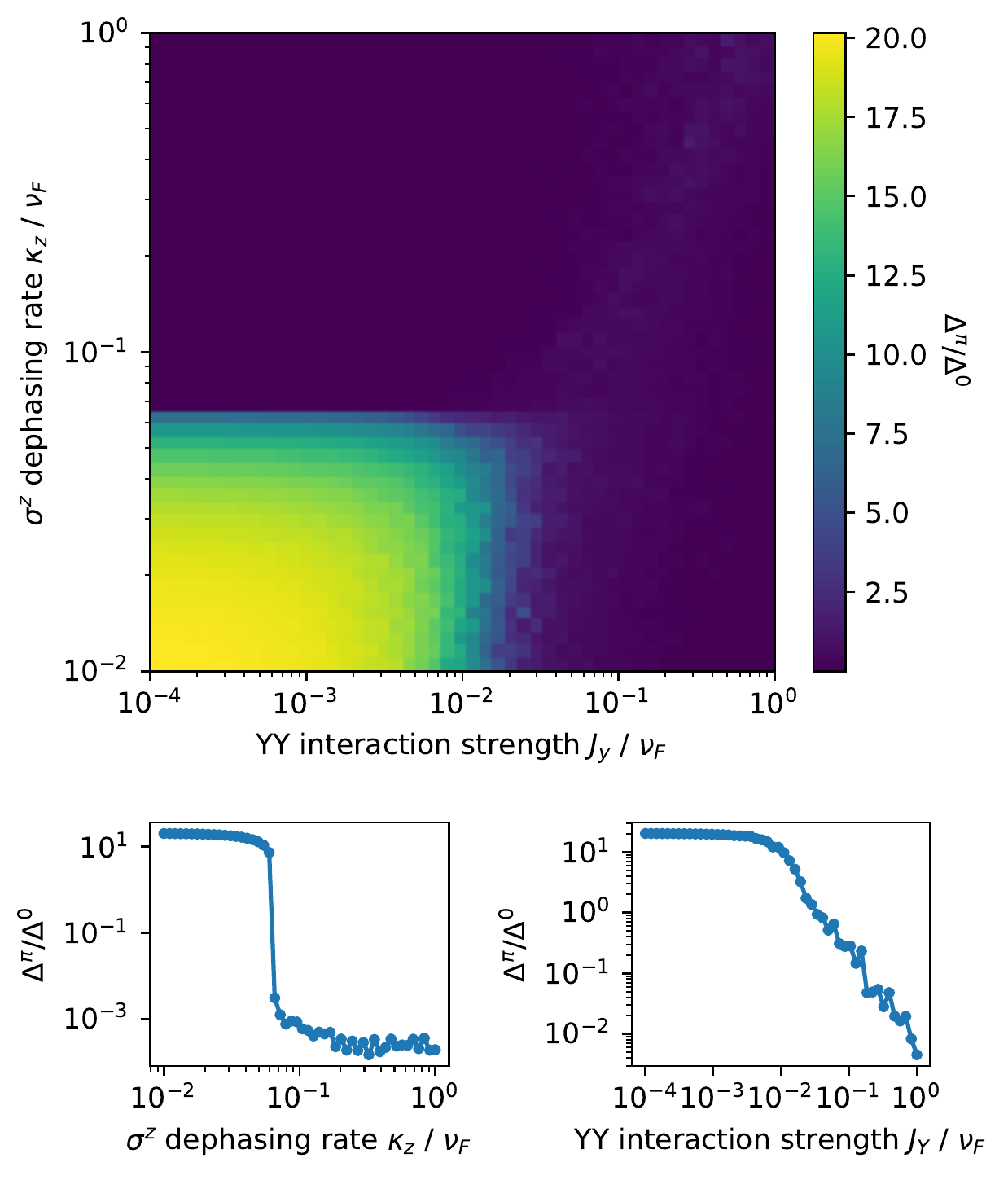}
        \label{fig:pi_gap_phase_diagram}
    }
    \caption{Phase diagram of the discrete time crystal with competing dissipation and interactions, using two different order parameters. \cref{fig:sigmaz_phase_diagram} includes experimental data exploring the strong dissipation/weak interaction phase boundary. \textbf{(a)} Phase diagram from the Fourier spectrum of $\langle\sigma^{z}_{0}\rangle$, calculating using the driven central spin model coupled to a dissipative bath. The DTC phase corresponds to the dark region, where $\langle \sigma_{0}^{z} \rangle$ has a stable peak at half the drive frequency. Unlike the $\pi$-gap phase diagram, there is no dark region in the bottom right because we are looking at $\sigma^{z}$; in this region $\sigma^{y}$ exhibits period doubling instead. The red and white stars and corresponding Fourier spectra correspond to experiments with $\tau_{\mathrm{lock}} = 10t_{\pi}$ and $\tau_{\mathrm{lock}} = 100t_{\pi}$ respectively. \textbf{(b)} $\pi$-gap phase diagram of the discrete time crystal as a function of dephasing rate and $\sigma^{y}_{i}\sigma^{y}_{j}$ interaction strength. The DTC phase corresponds to $\Delta_{\pi}/\Delta_{0} \ll 1$ (dark regions). The two plots at the bottom are slices through the phase diagram, the left at weak interactions and the right at weak dissipation. Both slices show a clear transition in the $\pi$-gap, though this is more pronounced for the dissipation-driven transition. \textbf{Parameters:} $\epsilon = 0.01$, $h = \SI{10}{\Hz}$, $N=5$, 10 disorder realizations; Fourier transforms calculated over $0 \leq n \leq 200$ Floquet periods. Phase diagram axes are in units of the Floquet frequency $\nu_{F}$.}
\end{figure*}

Discrete time crystals can be characterized by their robust subharmonic response to a periodic drive. 
Intuitively, the dissipation can cancel out any small perturbations to the state or the drive which might otherwise break the subharmonic response (see \cref{fig:bloch_floquet}). What is more non-trivial is whether this dissipation-stabilized subharmonic response can persist in the presence of destabilizing interactions. In this section we map out the phase diagram of a discrete time crystal (DTC) with competing dissipation and interactions, and numerically show that a dissipation-driven DTC is indeed stable to weak interactions. Further, we explore the opposite regime, where strong interactions stabilize the DTC, and show that this regime is stable to weak dissipation.

In the experiment outlined in \cref{sec:pulse_protocols,sec:experimental_observations}, we explore the strong dissipation/weak interaction regime of this phase diagram, and find that our observations align with the theoretically predicted phase boundary (see \cref{fig:sigmaz_phase_diagram}). At the other extreme, the weak dissipation/strong interaction regime could be probed by looking at samples with higher concentrations of phosphorous dopants. 

To map out the phase diagram of the dissipative DTC, we use two different probes which agree in their predictions for the DTC phase boundary. First, we use the experimentally-relevant probe of calculating the dynamics of local observables and whether or not they exhibit stable subharmonic peaks in their Fourier spectra (\cref{fig:sigmaz_phase_diagram}). Second, we look at the $\pi$-gap of the Floquet Liouvillian (\cref{fig:pi_gap_phase_diagram}), which we define in \cref{sec:order_parameters}. The $\pi$-gap was first introduced in Ref.\ \cite{vonkeyserlingkAbsoluteStabilitySpatiotemporal2016} to study emergent symmetries in many-body localized DTCs, but here we generalize it to Liouvillian systems.

\subsection{Theoretical model}
\label{sec:theoretical_model}

To produce the phase diagram, we focus on the driven central spin model (CSM) coupled to a dissipative bath (see \cref{fig:central_spin_model_cartoon}). The CSM has been successful as a semiclassical effective model for describing decoherence in solid state systems \cite{prokofevTheorySpinBath2000, Khaetskii2002_decoherence,  yuzbashyan2005CSM, Yao_2006_decoherenceQD, bortzExactDynamicsInhomogeneous2007, Chen_2007_CSQuantumDot, Cywinski2009_HF_dephasing, Barnes2012_CSM, chekhovichNuclearSpinEffects2013}. In the experiment described in \cref{sec:pulse_protocols,sec:experimental_observations}, the strongest coherent interaction is between the electronic spins of the phosphorous dopants, which form the spins in the CSM. Nuclear spins from residual \ce{{}^{29}Si} contribute to dissipation rates, and provide a small source of random field for the electron spins.

To have a notion of \textit{competing} interactions and dephasing, we use $\sigma^{y} \sigma^{y}$ interactions and $\sigma^{z}$ dephasing, which in principle should produce DTC-like phenomena in $\sigma^{y}$ and $\sigma^{z}$ respectively. Interestingly, in the weakly interacting case, we find that strong $\sigma^{z}$ dephasing can also stabilize a DTC-like response in $\sigma^{y}$.

The Hamiltonian of the central spin model with $\sigma^{y} \sigma^{y}$ interactions is given by

\begin{equation}
    H_{\mathrm{CSM}} = \sum_{i = 1}^{N} J_{0i} \sigma_{0}^{y} \sigma_{i}^{y} + \sum_{j = 0}^{N} h_{j} \sigma_{j}^{z},
    \label{eq:csm}
\end{equation}
where the central spin has index 0, and the outer spins have indices 1 to $N$. $J_{0i}$ and $h_{j}$ are modelled as random variables taken from the uniform distributions over $[-h, h]$ and $[-J, J]$ respectively. To incorporate the driving, the Hamiltonian follows a two-stage protocol given by 

\begin{equation}
    H(t) = \left\{\begin{alignedat}{2}
                & H_{\mathrm{CSM}}, && \text{for time }\tau_{\mathrm{lock}}; \\
                & (1 + \epsilon) \dfrac{\pi}{2 \tau_{\mathrm{flip}}} \sum_{i=0}^{N} \sigma_{i}^{x}, \hspace{1em} && \text{for time }\tau_{\mathrm{flip}}.
            \end{alignedat}\right.
\end{equation}
For $\epsilon = 0$, the second stage of the pulse protocol exactly flips the spins. The discrete time crystal phase can be defined operationally by its robustness against nonzero rotation errors $\epsilon$.

In addition to the interactions present within the central spin model, we model the effects of external dephasing using the Lindblad master equation given by

\begin{equation}
    \dfrac{\mathrm{d} \rho}{\mathrm{d} t} = -i \left[H(t), \rho\right] + \kappa_{1} \sum_{i = 0}^{N} \left( \sigma_{i}^{z} \rho {\sigma_{i}^{z}}^{\dagger} - \dfrac{1}{2} \left\{ {\sigma_{i}^{z}}^{\dagger} \sigma_{i}^{z} , \rho \right\} \right).
    \label{eq:master_equation}
\end{equation}

Note that this dissipation model enacts XY-dephasing, i.e.\ it draws the state of a single qubit to the $z$-axis of the Bloch sphere. This can be seen as being in competition with the $\sigma^{y}\sigma^{y}$ interactions present within the central spin model [\cref{eq:csm}].

In the experiment, $T_{1\rho}$ was measured to be \SI{193}{\micro\s} for the higher density sample, giving $\kappa_{1} \sim \SI{2.3}{\kilo\Hz}$, while $J$ and $h$ can be estimated from the phosphorous concentration and are of the order \SI{300}{\Hz} and \SI{10}{\Hz} respectively. Hence we expect XY-dephasing to dominate the dynamics of the experimental system.

In producing \cref{fig:3D_exp_vs_theory_comparison}, we also included $T_{1}$-type dissipation via the Lindblad operators $\sqrt{\kappa_{2}} \ket{0}_{i}\bra{1}_{i}$. This serves merely to slightly broaden the Fourier peaks, and does not significantly affect the phase boundary; hence it was neglected when producing the phase diagrams. Indeed, in the experiment, $T_{1}$ was measured to be \SI{1097}{\micro\s}, giving $\kappa_{2} \sim \SI{0.9}{\kilo\Hz}$. Hence the strongest factor affecting the dynamics of the spins is XY-dephasing, which is the mechanism that stabilizes the dissipative discrete time crystal.

\subsection{Order parameters for the DTC phase diagram}
\label{sec:order_parameters}

\subsubsection{Fourier spectra of local observables}

In an experiment, a DTC is typically detected by measuring the time-dependence of local observables (or averages thereof), and looking for oscillations at an integer fraction $1/n$ of the drive frequency $\nu_{0}$ which are robust against perturbations. This robustness can be detected by looking at the Fourier spectra of these local observables; the DTC phase will have a strong peak at $\nu_{0}/n$ which remains unsplit for a finite window of rotation errors. To use this to produce a phase diagram, we fix a nonzero rotation error $\epsilon$, and then for each set of parameters calculate the Fourier spectrum of $\langle\sigma^{z}_{0}\rangle$ in the driven-dissipative central spin model (see \cref{sec:theoretical_model}), assign a value of +1 if the peak is split and 0 if not, and average the result over 10 disorder realizations.  

In the strong dephasing limit, both the central and the outer spins exhibit robust period-doubling. The same is also true in the strong interactions limit, where interestingly the outer spins have Fourier spectra which split for \textit{larger} values of $\epsilon$ than the central spin.

Finally, in \cref{sec:scaling_analysis} we investigate the DTC lifetime as a function of system size. We estimate the lifetime of the DTC order using the width of the Lorentzian fit to the subharmonic peak. For the system sizes we have been able to access, we observe negligible dependence of the DTC lifetime on system size. This is true both in the weak interactions/strong dissipation limit, and the strong interactions/weak dissipation limit. Though our conclusions have to be tempered by finite-size limitations, these results indicate that the dissipative DTC order has a finite lifetime in the thermodynamic limit.

\subsubsection{$\pi$-gap}

Consider a periodically-driven, or `Floquet', system. If its dynamics are unitary, then time-evolution at integer multiples of the drive period $T$ can be generated using the \textit{Floquet unitary} $U_{F} = \mathcal{T} \exp( -i \int_{0}^{T} H(t) \mathrm{d}t /\hbar ) \equiv \exp( - i H_{F} T / \hbar)$, where $\mathcal{T}$ denotes time-ordering, and $H_{F}$ is an effective \textit{Floquet Hamiltonian}. If instead its dynamics are non-unitary, we can make an analogous statement: time-evolution at integer multiples of the drive period $T$ can be generated using the exponential of the \textit{Floquet Liouvillian} $\exp(\mathcal{L}_{F} T) \equiv \mathcal{T} \exp( \int_{0}^{T} \mathcal{L}(t) \mathrm{d}t)$, where $\mathcal{L}(t)$ is the Liouvillian at time $t$.

One can analyze the stroboscopic dynamics of the system by looking at the spectrum of the Floquet Liouvillian $\mathcal{L}_{F}$. The imaginary part of the spectrum describes the oscillatory modes, while the real part describes the decay modes. In the $\mathbb{Z}_{2}$ DTC phase, $\pi$\textit{-pairing} occurs, where the eigenvalues come in pairs with imaginary parts separated by $\pi$. Provided the initial state has significant overlap with these $\pi$-paired states, the subsequent dynamics will exhibit period doubling. The $\pi$-gap $\Delta_{\pi}$ provides a measure of the extent of $\pi$-pairing across the spectrum of $\mathcal{L}_{F}$, and is defined as
\begin{equation}
    \Delta_{\pi} = \mathbb{E}\left[ \left| \mathrm{Im}\left( \lambda_{i+\mathcal{N}/2} - \lambda_{i}\right) - \pi \right| \right].
    \label{eq:pi_gap}
\end{equation}
Here the $\lambda_{i}$ are the eigenvalues of $\mathcal{L}_{F}$, ordered according to their imaginary parts, $\mathcal{N}$ is the total number of eigenvalues, and the average $\mathbb{E}$ is taken uniformly across the spectrum of $\mathcal{L}_{F}$. For an order parameter, we look at $\Delta_{\pi}/\Delta_{0}$, where $\Delta_{0} = {\mathbb{E} \left[ \left| \mathrm{Im} \left( \lambda_{i+1} - \lambda_{i}\right) \right| \right]}$ is the mean nearest-neighbour spacing, which provides a relevant scale to compare against. The $\mathbb{Z}_{2}$ DTC phase corresponds to $\Delta_{\pi}/\Delta_{0} \ll 1$.

\subsection{Analysis of the phase diagram}

The $\pi$-gap has the advantage that it can probe the DTC transition for arbitrary observables. For our purposes, this means we can simultaneously analyze DTCs with i) strong dissipation and weak interactions, and ii) weak dissipation and strong interactions. \cref{fig:pi_gap_phase_diagram} shows that there are two clear transitions which occur: one at weak interactions upon increasing the dephasing strength, and one at weak dephasing upon increasing the interaction strength. Further, the positions of these transitions are stable against finite interactions or dissipation respectively.

The experiment described in \cref{sec:experimental_observations} probes the dephasing-driven DTC transition. The detail to \cref{fig:sigmaz_phase_diagram} shows experimental Fourier spectra for experiments with $\tau_{\mathrm{lock}} = 10t_{\pi}$ and $\tau_{\mathrm{lock}} = 100t_{\pi}$. Increasing $\tau_{\mathrm{lock}}$ lengthens the Floquet period and hence reduces the Floquet frequency~$\nu_{F}$. Since the axes in the phase diagrams are measured in units of $\nu_{F}$, this allows us to explore regions of the phase diagram with effectively stronger interactions and dephasing. These experimental Fourier spectra thus probe either side of the dissipative DTC phase boundary, and demonstrate that the dissipative DTC is stable in the predicted region, even with potentially destabilizing interactions.



\section{Conclusions}

We have shown robust experimental signatures of the formation of discrete time-crystal phase in naturally purified silicon doped with phosphorus atoms, using composite BB1 pulses for exceptionally uniform rotation of the spin ensemble. We observe the formation of discrete time-crystalline order by driving the electron spin ensemble at frequency $\nu_0$, while producing a response at a sub-harmonic frequency $\nu_0/2$. We show that this peak remains pinned and is robust to perturbations in the periodic pulse protocol. Motivated by the experimental system, we investigate the dissipative central spin model as a phenomenological description for time-crystalline behaviour in solid state systems with long-range interactions. We show that the model, with no free parameters, is in remarkably good agreement with experiment. Furthermore, we investigate the role of interactions and dissipation stabilizing the DTC phase in the central spin model and the crossover regime in which both effects are significant.

The high level of quantum control of electron and nuclear spins in phosphorus doped silicon provides a promising platform for studying many-body quantum coherence in driven systems. By driving the nuclear and electronic spins independently, proximity effects in time-crystalline behaviour and effects of dynamic nuclear polarization can be explored. The difference in the dynamic time scales of electron and nuclear spins could serve as a useful tool for manipulating DTC order and its long time coherence. Furthermore, the life time of the time crystal order can be exploited to probe the dephasing and thermalizing properties of long range systems. The DTC order exhibited in the central spin model poses several interesting questions on non-thermal steady states in Floquet systems which are ripe for further investigation \cite{Hone_FloquetStatMech_2009, Haldar_FloquetThermal_2018}.

\section{Acknowledgements}
We acknowledge useful discussions with Dmitry Abanin, Fabian Essler, Masaki Oshikawa, Shivaji Sondhi, and Marzena Szymanska. At the time of the preparation of the manuscript, we came across some related complementary works studying the interplay between dissipation and DTC order \cite{choi2018_DTC_Thermal, gambettaDiscreteTimeCrystals2019}. 
J.O'S.\ thanks the EPSRC for a Doctoral Training Grant and J.J.L.M acknowledges funding from the European Research Council under the European Union's Horizon 2020 research and innovation programme (grant agreement No. 771493 (LOQO-MOTIONS). A.P.\ thanks the hospitality at the Erwin Schroedinger Institute, Vienna where part of the work was completed. This work was supported by the Engineering and Physical Sciences Research Council [grant number EP/L015242/1]. J. O. S. and O. L. have contributed equally to this work.

\appendix
\counterwithin{figure}{section}
\section{Effects of inhomogeneous spin rotations on the crystalline fraction}
\begin{figure*}[htp]
	\centering
	\subfloat[]{
		\centering
		\includegraphics[width=0.45\linewidth]
		{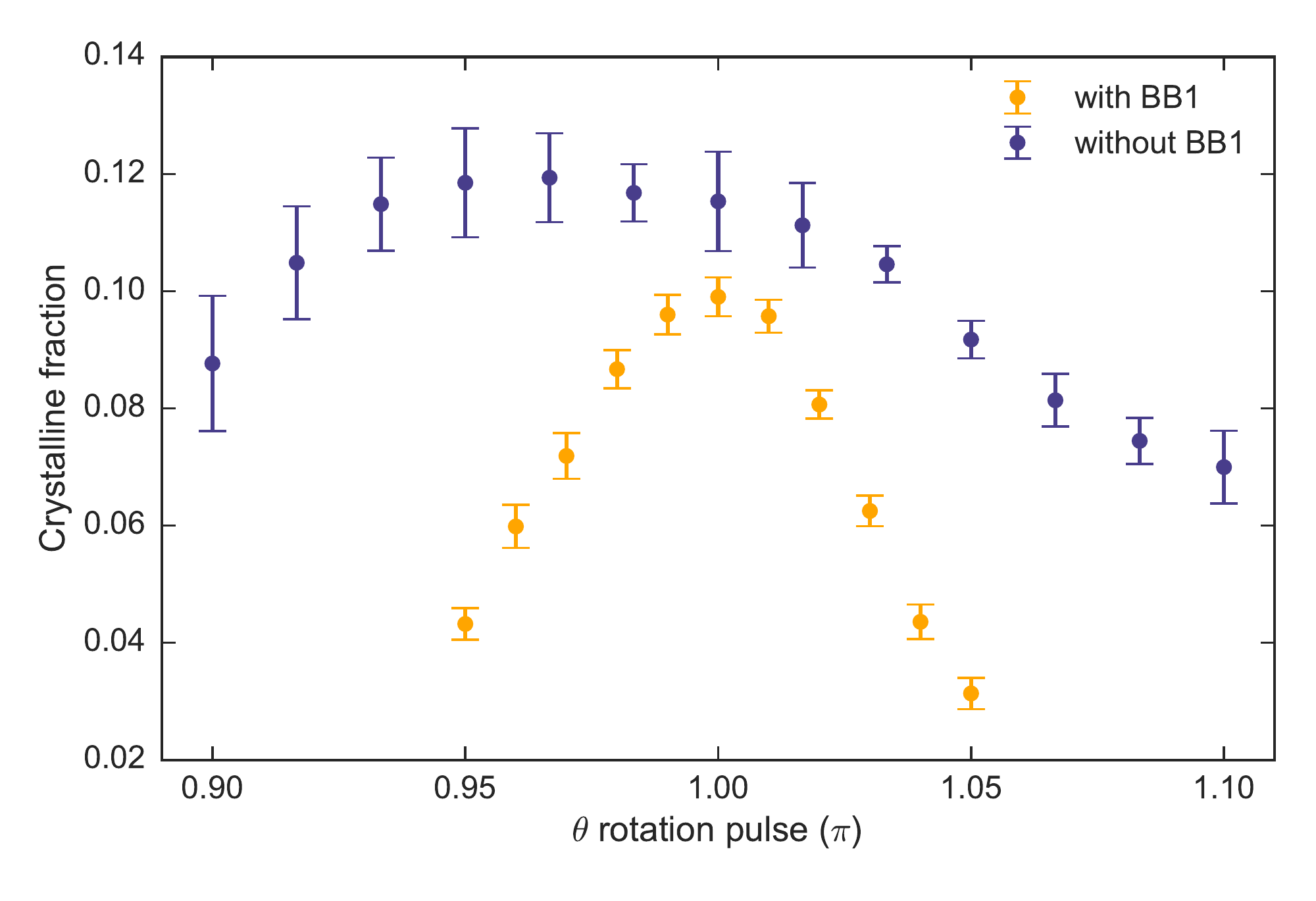}
		\label{fig:experimental_plateau}
	}
	\subfloat[]{
		\centering
		\includegraphics[width=0.45\linewidth]{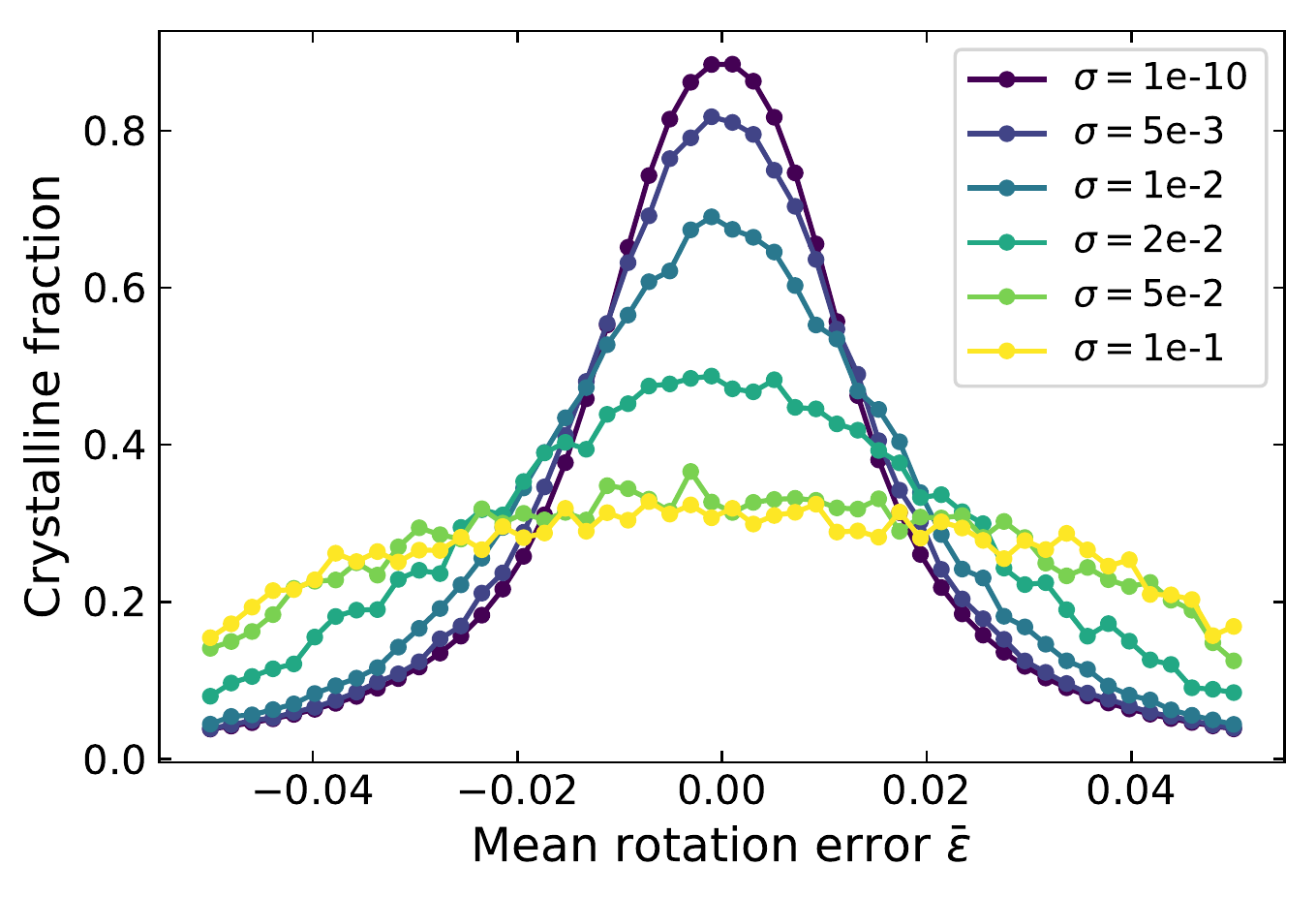}
		\label{fig:simulation_plateau}
	}
	\caption{The effect of nonuniform rotations on the crystalline fraction. \textbf{(a)} Comparison of the crystalline fraction against rotation error of the same sample with and without BB1 pulses turned on. The non-BB1 pulse sequence appears to have a somewhat flattened peak, while the BB1 crystalline fraction is a peaked curve. The subharmonic oscillations of the Floquet cycles without BB1 also decayed much faster than the BB1 oscillations. A longer $\pi$ duration of 300ns in the BB1 data vs 200ns for the non-BB1 data, causing more dephasing and weaker signal, may account for the lower overall crystalline fraction of the BB1 experiment. \textbf{(b)} Simulations of the crystalline fraction using the driven-dissipative central spin model using nonuniform rotation pulses. For a given disorder realization, the rotation error $\epsilon$ is drawn from a Gaussian distribution with mean $\bar{\epsilon}$ and variance $\sigma^{2}$, clipped at 10\% to model the finite extent of the sample. We observe that large values of $\sigma$ result in a distinct flattening of the crystalline fraction curve, consistent with our experimental observations in \cref{fig:experimental_plateau}.}
	\label{fig:plateau}
\end{figure*} 
\label{sec:BB1}


The crystalline fraction, defined in terms of the Fourier transform $S(\nu)$ as $|S(\nu=1/2)|^{2}/\sum_{\nu} |S(\nu)|^{2}$, has been recently utilised as an experimental probe of $\mathbb{Z}_{2}$ discrete time crystal robustness \cite{Choi2017,choi2018_DTC_Thermal,Rovny_DTC2018}. A `plateau' in the crystalline fraction in a finite window around zero rotation error has been used an indicator for DTC order. However, we have observed that this plateau can emerge simply as a result of having nonuniform rotation pulses across the extent of the sample.

Experimentally, this nonuniform rotation comes in our case from variations in the microwave frequency magnetic field generated by the split ring cavity used for driving and detection. To facilitate a comparison of the effect of these nonuniform rotations, we used a composite rotation pulse known as a BB1 pulse, which allows us to correct variations across our sample, estimated to be on the order of $\pm10\%$, up to 5\textsuperscript{th} order. The BB1 pulse takes the form of a simple $\theta$ rotation pulse, immediately followed by a $\pi(\phi)-2\pi(3\phi)-\pi(\phi)$ corrective pulse sequence, where the phase of each pulse $\phi = \arccos(-{\theta}/{4\pi})$. We used a BB1 pulse during the main experiment to ensure an extremely uniform rotation of the spin ensemble, which enables us to properly diagnose the robustness of the DTC phase to deliberate rotation errors.

\cref{fig:experimental_plateau} shows a comparison of the crystalline fraction with and without BB1 pulses. It should be noted that while the two experiments were essentially the same, the experimental apparatus used was different between these two runs. Without BB1, the nonuniform rotations result in a flattening of the crystalline fraction, which could be seen as a `false-positive' for DTC order. However, if we do use BB1 pulses to ensure uniform rotations, the `plateau' disappears, and the crystalline fraction is simply peaked around the point of zero pulse error. To corroborate these results, we performed simulations of the driven-dissipative central spin model using nonuniform rotation pulses (\cref{fig:simulation_plateau}). For a given disorder realization, the rotation error $\epsilon$ is drawn from a Gaussian distribution with mean $\bar{\epsilon}$ and variance $\sigma^{2}$. The maximum deviation from $\bar{\epsilon}$ is fixed at 10\% to model the finite extent of the sample. The same rotation error is used periodically up to 200 Floquet cycles, from which we calculate the crystalline fraction. This is repeated 500 times and averaged to produce a given curve in \cref{fig:simulation_plateau}. Here we observe results consistent with the experiment: larger variance $\sigma^{2}$ in the rotation error results in a distinct flattening of the crystalline fraction.

\section{Derivation of $\epsilon^{2}$ peak width dependence}
\label{sec:derivation}
\begin{figure}[t]
	\includegraphics[width=\linewidth]{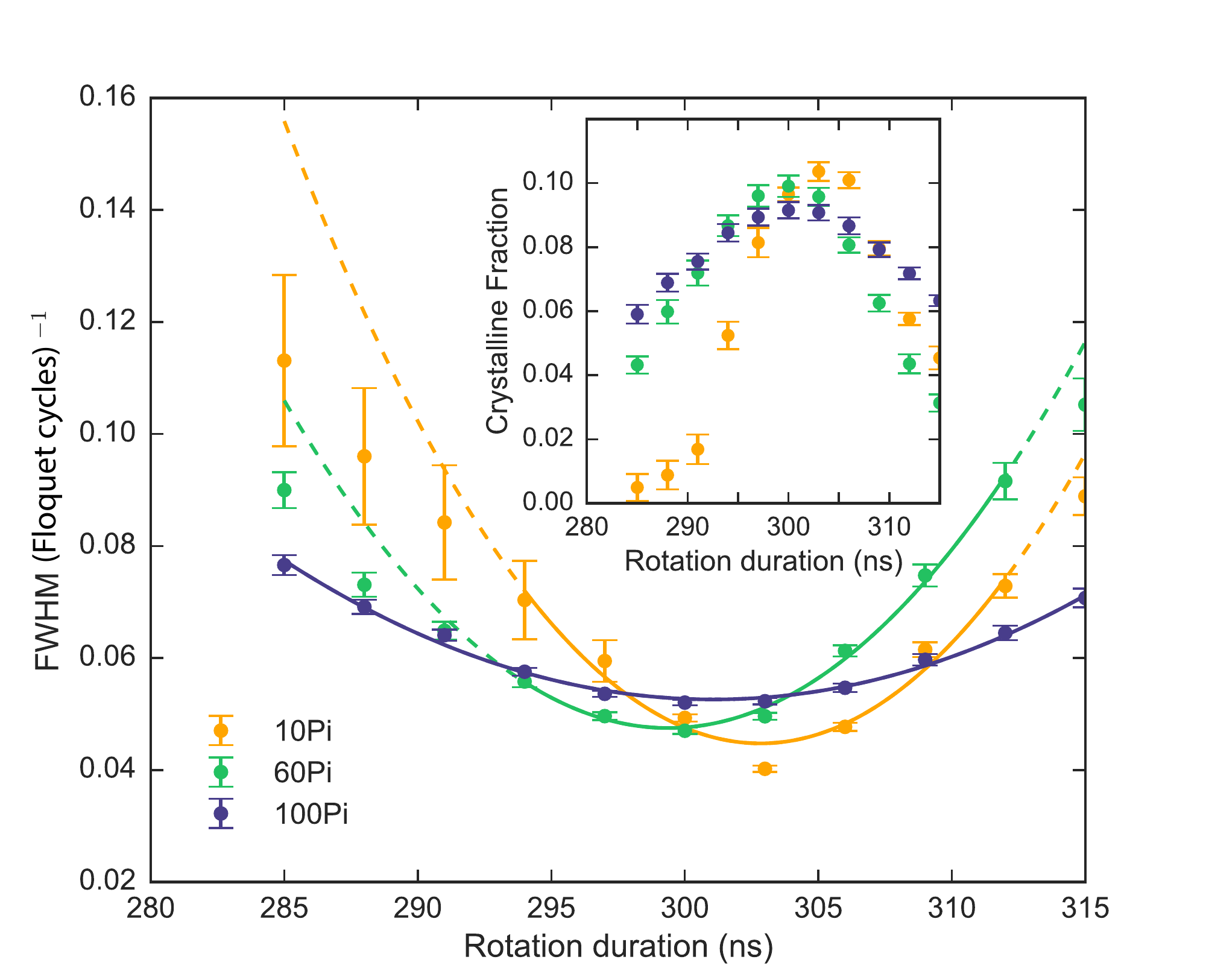}
	\caption{Peak width and crystalline fraction as a function of rotation duration for the higher concentration sample, doped to \SI{3e15}{\cm^{-3}}. We can see the key features are essentially the same as \cref{fig:experiment_peak_width}, indicating that we are still in the dephasing dominated regime at this concentration.}
	\label{fig:high_conc}
\end{figure} 
\begin{figure*}[htp]
    \centering
    \subfloat[]{
        \centering
        \includegraphics[width=0.45\linewidth]{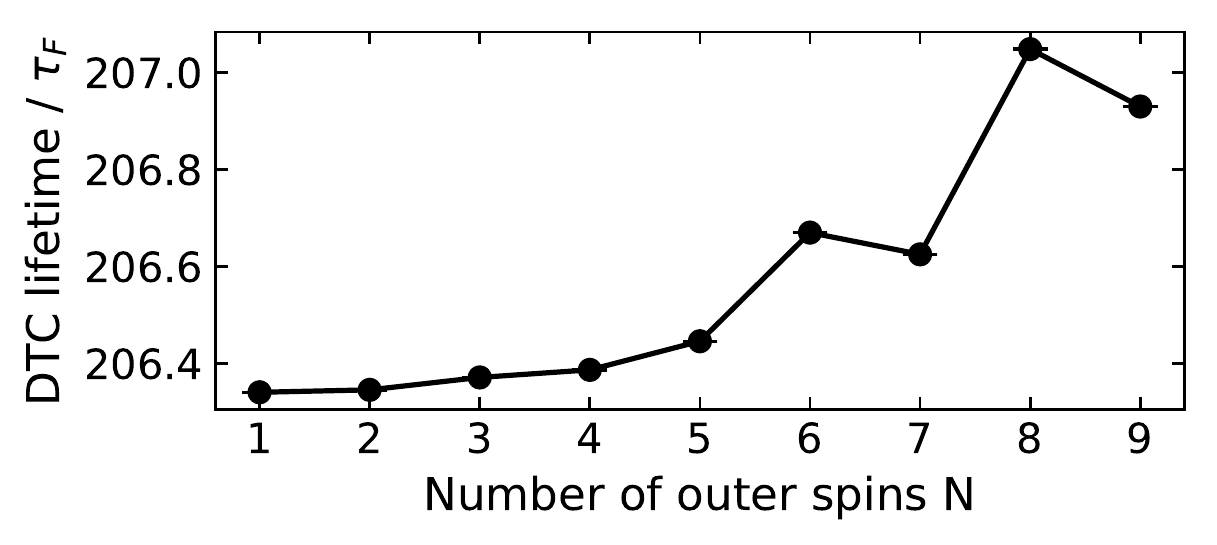}
        \label{fig:DTC_lifetime_strong_dephasing}
    }
    \subfloat[]{
        \centering
        \includegraphics[width=0.45\linewidth]{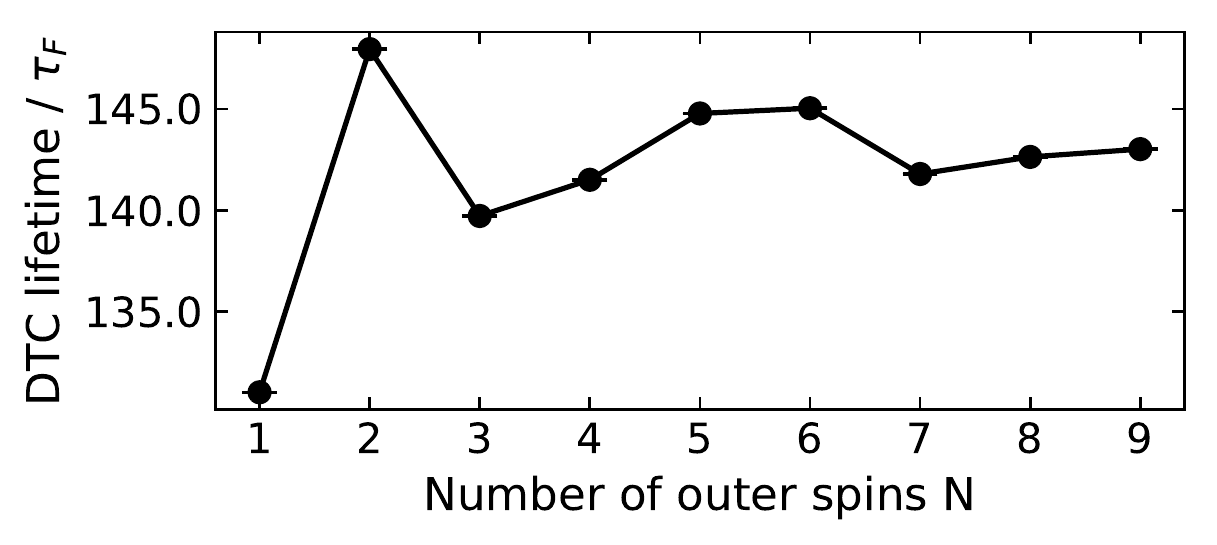}
        \label{fig:DTC_lifetime_weak_dephasing}
    }
    
\caption{Lifetime of the discrete time crystal (DTC) as a function of system size. Both plots were generated by simulating the dynamics of local observables under the driven central-spin model coupled to a dissipative bath, and fitting a Lorentzian to the subharmonic peak in their Fourier spectra. The inverse of the full-width-at-half-maximum of the Lorentzian gives the DTC lifetime. The parameters for \textbf{(a)} and \textbf{(b)} correspond to the top-left and bottom-right corners of \cref{fig:sigmaz_phase_diagram} respectively. \textbf{(a)} Strong dephasing and weak interactions: dephasing rate $\kappa = \nu_{F}$ and $\sigma^{y}\sigma^{y}$ interaction strength $J = 10^{-4} \nu_{F}$. The Fourier spectrum of $\sigma^{z}_{0}$ was used to calculate these lifetimes. \textbf{(b)} Weak dephasing and strong interactions: dephasing rate $\kappa = 10^{-2} \nu_{F}$ and $\sigma^{y}\sigma^{y}$ interaction strength $J = \nu_{F}$. The Fourier spectrum of $\sigma^{y}_{0}$ was used to calculate these lifetimes. \textbf{Parameters:} $\epsilon = 0.01$, $h = \SI{10}{\Hz}$, 10 disorder realizations; Fourier transforms calculated over $0 \leq n \leq 200$ Floquet periods. Lifetimes are in units of the Floquet period $\tau_{F}$, and frequencies are in units of the Floquet frequency $\nu_{F} = 1/\tau_{F}$.}

\label{fig:DTC_lifetimes}

\end{figure*} 
In the strong-dephasing limit, many of the qualitative features of a dissipative discrete time crystal can be understood by analyzing a simple model consisting of a single spin exposed to external dephasing. We assume that the rotation phase of the Hamiltonian pulse protocol is performed instantaneously. In practice this is not quite true, and the action of dephasing during a finite rotation time can lead to further broadening of the Fourier peak, but this broadening is independent of $\epsilon$, so here we neglect it for simplicity. We also assume that we are in the limit where $T_{2} \ll T_{1}$ so that we can neglect $T_{1}$-type relaxation. A similar derivation appears in Ref.\ \cite{choi2018_DTC_Thermal}.

We use the Liouvillian formalism of open quantum systems to perform the derivation. Recall that the Liouvillian $\mathcal{L}$ is defined by $\frac{\mathrm{d} \rho}{\mathrm{d} t} = \mathcal{L} \rho$, which for time-independent $\mathcal{L}$ has the solution $\rho(t) = \exp(\mathcal{L} t) \rho(0)$. The product $\mathcal{L} \rho$ should be understood as a matrix-vector product, where $\rho = (\rho_{00}, \rho_{01}, \rho_{10}, \rho_{11})^{\mathrm{T}}$ and $\mathcal{L}$ is a $4 \times 4$ matrix.

Following \cref{eq:master_equation}, the XY-dephasing can be implemented by the Liouvillian
\begin{equation}
    \mathcal{L}_{\mathrm{XY}} = -\kappa\Bigl( \oouter{01} + \oouter{10} \Bigr),
\end{equation}
where $\kappa$ is the XY-dephasing rate, and we use the convention $\rho = \sum_{ij} \rho_{ij} \ket{i}\bra{j} \mapsto \kket{\rho} = \sum_{ij} \rho_{ij} \ket{i}\otimes\ket{j}$. The spin-flip is implemented by the unitary $U_{\mathrm{flip}} = \exp(-i \sigma^{x}(1 + \epsilon) \pi/2 )$, which is equivalent to the Liouvillian $\exp(\mathcal{L}_{\mathrm{flip}}) = U_{\mathrm{flip}} \otimes U_{\mathrm{flip}}^{\dagger}$. The Liouvillian which governs time-evolution over one Floquet period, termed the Floquet Liouvillian $\mathcal{L}_{F}$, is then given by $\exp(\mathcal{L}_{F} \tau) = \exp(\mathcal{L}_{\mathrm{flip}}) \exp(\mathcal{L}_{\mathrm{XY}} \tau)$, where $\tau$ is the Floquet period.

The density matrix after $n$ Floquet cycles can then be calculated as $\rho(n\tau) = (e^{\mathcal{L}_{F} \tau})^{n} \rho(0)$. For simplicity, we consider only the limit $\kappa \tau \gg 1$. In this limit, $e^{\mathcal{L}_{F} \tau}$ has eigenvalues $\lambda \in \{1,0,0,-\cos(\epsilon\pi)\}$ with corresponding eigenvectors
\begin{equation}
    \rho_{\lambda} = \begin{matrix}
        \dfrac{1}{2}\begin{pmatrix}1\\0\\0\\1\end{pmatrix}, & \begin{pmatrix}0\\1\\0\\0\end{pmatrix}, & \begin{pmatrix}0\\0\\1\\0\end{pmatrix}, & \dfrac{1}{2}\begin{pmatrix}1\\i\tan(\epsilon\pi)\\-i\tan(\epsilon\pi)\\-1\end{pmatrix}.
    \end{matrix}
\end{equation}

Note that for $\epsilon \neq 0$, the unique steady state is the maximally mixed state $\frac{1}{2}(1,0,0,1)^{\mathrm{T}}$, which is a Floquet eigenvector, so we can deduce that the dissipative DTC has a finite lifetime for any nonzero $\epsilon$, even in the limit of infinite dephasing.

Choosing the initial state $\rho(0) = |0\rangle\langle 0|$ and expanding this in the basis of eigenvectors of the Floquet Liouvillian, we find that the state after $n$ Floquet periods is
\begin{equation}
    \rho(n\tau) = \dfrac{1}{2}\begin{pmatrix}1\\0\\0\\1\end{pmatrix} + \dfrac{1}{2}\left[-\cos(\epsilon\pi)\right]^{n} \begin{pmatrix}1\\i\tan(\epsilon\pi)\\-i\tan(\epsilon\pi)\\-1\end{pmatrix},
\end{equation}
and using $\langle \sigma^{z} \rangle = \mathrm{Tr}\left[\sigma^{z} \rho \right] = \rho_{00} - \rho_{11}$, we have
\begin{equation}
    \langle \sigma^{z} \rangle(n\tau) = [-\cos(\epsilon\pi)]^{n}.
\end{equation}
Hence $|\langle\sigma^{z}\rangle|$ decays exponentially at the rate $\Gamma \tau = -\log[\cos\epsilon\pi] \sim (\epsilon\pi)^{2}/2$ per Floquet period. 

\section{Scaling analysis}
\label{sec:scaling_analysis}

In this section we analyze how the lifetime of the discrete time crystal changes with system size. We will explore this question in two regimes: strong dephasing and weak interactions, and weak dephasing and strong interactions. In both cases, we extract the lifetime by calculating the Fourier spectrum of a local observable: $\sigma^{z}$ for strong dephasing, and $\sigma^{y}$ for strong interactions. From \cref{fig:sigmaz_phase_diagram} and the analogous diagram for $\sigma^{y}$ (not shown), we know that the DTC is stable in these limits. This ensures that we can reliably fit a single Lorentzian to the Fourier spectrum, and from there extract the lifetime as the inverse of the full-width-at-half-maximum of the Lorentzian.

\cref{fig:DTC_lifetime_strong_dephasing,fig:DTC_lifetime_weak_dephasing} show the DTC lifetime as a function of system size in the case of strong dephasing and weak interactions, and weak dephasing and strong interactions respectively. In both cases, we observe negligible dependence of the lifetime on system size, though the degree to which we can make this conclusion is limited by the system sizes accessible by numerically exact dynamics. We note that finite lifetimes in the thermodynamic limit have been reported in other dissipative DTC systems \cite{gambettaDiscreteTimeCrystals2019}; in general one should not expect these dissipative systems to have infinite lifetimes in the thermodynamic limit.

%

\newpage
\bibliography{Time_Crystal}

\end{document}